\documentclass[12pt]{article}

\usepackage{scicite}

\usepackage{times}

\usepackage{graphicx}
\usepackage{bm}

\newenvironment{sciabstract}{%
\begin{quote} \bf}
{\end{quote}}

\newcommand{\onlinecite}[1]{\hspace{-1 ex} \nocite{#1}\citenum{#1}}

\newcounter{lastnote}

\title{Critical enhancement of thermopower in a chemically tuned polar semimetal MoTe$_{\bf 2}$} 
\author
{Hideaki Sakai$^{1,2\dagger}$, Koji Ikeura$^1$, Mohammad Saeed Bahramy$^{3,4}$, Naoki Ogawa$^{4}$,\\ Daisuke Hashizume$^{4}$, Jun Fujioka$^{1,5}$, Yoshinori Tokura$^{1,4}$, Shintaro Ishiwata$^{1,5*}$\\
\\
\normalsize{$^{1}$Department of Applied Physics, University of Tokyo, Tokyo 113-8656, Japan}\\
\normalsize{$^{2}$Department of Physics, Osaka University, Toyonaka, Osaka 560-0043, Japan}\\
\normalsize{$^{3}$Quantum-Phase Electronics Center (QPEC) and Department of Applied Physics,}\\
\normalsize{ University of Tokyo, Tokyo 113-8656, Japan}\\
\normalsize{$^{4}$RIKEN Center for Emergent Matter Science (CEMS), Wako, Saitama 351-0198, Japan}\\
\normalsize{$^{5}$PRESTO, Japan Science and Technology Agency, Kawaguchi, Saitama 332-0012, Japan}\\ \\
\normalsize{$^\dagger$E-mail: sakai@phys.sci.osaka-u.ac.jp}\\
\normalsize{$^\ast$E-mail: ishiwata@ap.t.u-tokyo.ac.jp}
}

\date{}

\begin{document} 


\baselineskip24pt


\maketitle 

%
{\bf [Summary sentence]\\}
Unusual enhancement of cryogenic thermopower manifests itself around the critical point of polar order in a metal.
%
%
\begin{sciabstract}
Ferroelectrics with spontaneous electric polarization play an essential role in today's device engineering, such as capacitors and memories.
Their physical properties are further enriched by suppressing the long-range polar order, as is exemplified by quantum paraelectrics with giant piezoelectric and dielectric responses at low temperatures.
Likewise in metals, a polar lattice distortion has been theoretically predicted to give rise to various unusual physical properties.
So far, however, a ``ferroelectric''-like transition in metals has seldom been controlled and hence its possible impacts on transport phenomena remain unexplored.
Here we report the discovery of anomalous enhancement of thermopower near the critical region between the polar and nonpolar metallic phases in 1T'-Mo$_{1-x}$Nb$_{x}$Te$_2$ with a chemically tunable polar transition.
It is unveiled from the first-principles calculations and magnetotransport measurements that charge transport with strongly energy-dependent scattering rate critically evolves towards the boundary to the nonpolar phase, resulting in large cryogenic thermopower.
Such a significant influence of the structural instability on transport phenomena might arise from the fluctuating or heterogeneous polar metallic states, which would pave a novel route to improving thermoelectric efficiency.
\end{sciabstract}
%
%
\section*{Introduction}
%
A ``ferroelectric''-like transition in metals was first predicted by a pioneering work\cite{Anderson1965PRL} in 1960's, followed by the predictions of fascinating physical properties of a polar metal, such as unconventional superconductivity\cite{Saxena2004Nature}, magneto-optical effects\cite{Edelstein1998PRL}, and highly anisotropic thermopower\cite{Puggioni2014Natcom}.
At present, however, there are few reported metallic materials showing a polar structural transition\cite{Shi2013NatMat,Liu2015PRB,Sergienko2004PRL,Kolodiazhnyi2010PRL,Fujioka2015SR}.
In particular, control of the structural transition temperature has remained as an experimental challenge, which would enable the search for novel quantum phenomena associated with the criticality of the polar, i.e., ``ferroelectric''-like, transition in metals.
%
\par
%
To seek an ideal metallic system with a chemically tunable polar structural transition, we have here focused on transition metal dichalcogenides (TMDs).
TMDs have attracted renewed interest because of a rich variety of electronic properties, which are associated with large structural variations ranging from chemically controllable bulk forms to mechanically exfoliated single (or a few) layers\cite{Wang2012NatNano, Chhowalla2013NatChem}.
Among metallic TMDs, 1T'-$M$Te$_2$ [e.g. $M$ = Mo (Refs.\onlinecite{Vellinga1970JSSC, Hughes1978JPhysC, Zandt2007JAlloy, Ikeura2015APLMat}) and W (Ref.\onlinecite{Kabashima1966JPSJ})] has gained a great attention, spurred by the discoveries of extremely large magnetoresistance effects\cite{Ali2014Nature, Keum2015NatPhys} and pressure-induced superconductivity\cite{Kan2015NatCom, Pan2015NatCom, Qi2015Natcom}, and the prediction of Weyl semimetallic\cite{Soluyanov2015Nature, Sun2015PRB} and quantum Hall insulating states\cite{Qian2014Science}.
1T'-$M$Te$_2$ crystallizes in CdI$_2$-type structure consisting of edge-sharing $M$Te$_6$ octahedra with a strong distortion, which is caused by the formation of zigzag chains of metal-metal bonding along the $b$ axis\cite{Brown1966Acta} (as enclosed by dashed curves in Fig. 1A).
At room temperature, however, the crystal structures for 1T'-MoTe$_2$ and 1T'-WTe$_2$ differ from each other\cite{Brown1966Acta,Dawson1987JPCSSP}; the former is monoclinic ($P2_1/m$, Fig. 1A right) and the latter is orthorhombic ($Pnm2_1$, Fig. 1A left).
The point to be noted here is that 1T'-WTe$_2$ has a polar (noncentrosymmetric) space group, which results in the Weyl semimetallic state\cite{Chang2016NatCom}.
For 1T'-MoTe$_2$, on the other hand, it was reported that the centrosymmetric monoclinic structure at room temperature changes to orthorhombic structure at 250 K\cite{Clarke1978PhilMag, Manolikas1979PSSA}, which may be identical to the one for 1T'-WTe$_2$.
Therefore, 1T'-MoTe$_2$ is a rare and promising candidate of the metallic system allowing chemical control of a polar structural transition. 
%
\section*{Results}
%
\subsection*{Chemically-tunable polar transition in a metal.}
%
To demonstrate the tuning of the putative polar structural transition at $\sim$250 K in 1T'-MoTe$_2$, we prepared single crystalline samples of Mo$_{1-x}$Nb$_x$Te$_2$ (for determination of the chemical composition, see fig. S1 and Methods) and performed the x-ray diffraction measurements at selected temperatures.
The panels in inset to Fig. 1B present the CCD images of Bragg reflections (1 0 13 and \={1} 0 13) sensitive to the symmetry change, where the horizontal direction corresponds to the Bragg angle.
Upon cooling for $x=0$, the two reflections that clearly split up by the monoclinic distortion at 300 K approach to each other at 250 K and coalesce to be a single 1 0 13 reflection below 200 K, indicating a transition to an orthorhombic phase\cite{Clarke1978PhilMag, Manolikas1979PSSA}.
A single-crystal x-ray structural analysis at 100 K has indeed revealed that the space group of the low-temperature phase is orthorhombic $Pnm2_1$, where the polar direction is along the $c$ axis (Fig. 1A and see table S1 for details).
To be emphasized here is that the polar transition temperature in metallic 1T'-MoTe$_2$ can be systematically tuned from $\sim$250 K down to zero by chemical substitution of Nb for Mo.
As shown in inset to Fig. 1B, for $x=0.08$, the temperature below which two split Bragg reflections coalesce is lowered by more than 50 K compared to the pristine $x=0$ compound.
For $x=0.22$, the structural transition is further suppressed; the Bragg spots remain to be split even at 100 K.
The suppression of the polar transition can be associated with a monotonic decrease in interlayer distance upon Nb substitution\cite{Ikeura2015APLMat}, since the interlayer distance unusually increases with decreasing temperature across the polar transition\cite{Clarke1978PhilMag}.
This is apparently consistent with the fact that the external pressure reduces the polar transition temperature for 1T'-MoTe$_2$\cite{Kan2015NatCom, Pan2015NatCom, Qi2015Natcom}.
%
\par
%
The resultant structural phase diagram is shown in Fig. 1B, where the transition temperature $T_{\rm S}$ is precisely determined by the transport and optical measurements ($vide$ $infra$).
The value of $T_{\rm S}$ gradually decreases with increasing $x$, followed by an immediate drop to zero at around $x=0.1$.
For $x=0.12$, the polar orthorhombic phase is replaced by the nonpolar monoclinic one at the ground state, forming a critical state between the polar and nonpolar metals.
Note here that such a controllable polar transition is quite rare in metallic systems, considering that even in degenerately-doped ferroelectrics, such as oxygen-deficient and La-doped BaTiO$_{3}$\cite{Kolodiazhnyi2010PRL,Fujioka2015SR}, the ``ferroelectric'' transition temperature remains nearly constant (above 250-300 K) upon carrier doping.
Below we shall present a marked impact of tuning the polar transition on the transport and optical properties.

%
\par
%
Figure 2A shows the temperature profile of resistivity for 1T'-Mo$_{1-x}$Nb$_{x}$Te$_2$ with $x$ from 0 to 0.22.
The $x=0$ compound exhibits good metal behavior with the residual resistivity $\rho_0\!=\!5.9$ $\mu\Omega$cm at 2 K, resulting in the ratio of the room temperature resistance to that at 2 K (RRR) $\sim$65.
A clear resistivity anomaly at $T_{\rm S}\!\sim\!250$ K with a thermal hysteresis (denoted by arrows) reflects the first-order nature of the polar structural transition.
As $x$ increases to 0.08 through 0.03, the values of $T_{\rm S}$ defined by the resistivity anomalies decrease, accompanied by large thermal hysteresis between the cooling and heating runs.
For $x\!\ge\!0.12$, no anomaly is observed down to the lowest temperature, indicating that the polar structural transition is completely suppressed.
Interestingly, the variation of $\rho_{0}$ is nonmonotonic as a function of $x$; it rapidly increases from 5.9 to 177 $\mu\Omega$cm with increasing $x$ from 0 to 0.08, followed by a decrease by about half for $x=0.12$.
For $x\!\ge\!0.12$, $\rho_0$ is almost independent of $x$.
The anomalous $x$ dependence of $\rho_{0}$ cannot be simply explained by the impurity scattering due to the doped Nb ions, but may be relevant to the critical enhancement of charge scattering around $x=0.08$, as discussed below.
%
\par
%
Figure 2B presents the temperature profile of optical second harmonic generation (SHG) intensity for the samples with selected compositions.
Strong SHG signals are observed for $x=0$ and 0.08 below $T_{\rm S}$ but not for $x=0.22$, which supports the absence and presence of inversion symmetry in the orthorhombic and monoclinic phase, respectively.
The detailed SH polarization patterns are consistent with the point group symmetry $mm2$ inferred from the crystal structural analysis (figs. S2C and S2D) and the recent Raman spectroscopy\cite{Chen2016NanoLett,Zhang2016arxiv}.
The estimated nonlinear optical susceptibility is comparable in magnitude to the $d_{11}$ of quartz (see Supplementary Materials).
Note here that the temperature dependence of the SHG intensity for $x=0.08$ markedly differs from that for $x=0$.
For $x=0.08$ with a lower $T_{\rm S}$, the transition accompanied by a significant thermal hysteresis appears to be more diffusive; the nonzero SHG intensity remains even above $T_{\rm S}$.
Furthermore, the SHG intensity decreases with decreasing temperature below $\sim$150 K, which signals that the polar phase tends to be reduced in volume fraction, or fluctuated at low temperatures, since the Nb concentration is close to the critical value at the phase boundary ($x\!\sim\!0.1$).
%
\par
%
On the basis of the rigid-band scheme, the Nb substitution for Mo should decrease the Fermi energy, leading to the hole carrier doping.
This tendency is confirmed in the $x$ variation of Hall resistivity $\rho_{yx}$, as shown in Fig. 3B.
For $x=0$, the field profile of $\rho_{yx}$ is curved with a negative slope and strongly temperature dependent, which is typical of a semimetal with multi-carriers with opposite polarity.
This is consistent with the results of first-principles calculations, which predicts the formation of small electron and hole pockets (Fig. 4D, $x=0$), as is the case for WTe$_2$\cite{Ali2014Nature}.
Note here that a large magnetoresistance effect was also observed ($\sim$300\% at 2 K at 9 T) for $x=0$ (inset to Fig. 3B).
For $x=0.08$ and 0.22, on the other hand, the field profile of $\rho_{yx}$ becomes straight with positive slope for all the temperatures and the magnetoresistance effect is largely suppressed.
This indicates that only hole-like carriers exist for $x\!\ge\!0.08$, as schematically described in Fig. 4D ($x=0.1$ and $0.2$).
The Hall coefficient is almost temperature independent for $x=0.22$, as is common for simple metals.
For $x=0.08$, however, it exhibits an unusual temperature dependence; the value almost triples as the temperature decreases from 300 K to 2 K.
%
\subsection*{Thermopower enhancement near the critical region.}
%
The most dramatic impact of the suppression of the polar transition in 1T'-MoTe$_2$ is found in thermopower $S$ at low temperatures (Fig. 3A).
For $x=0$, the value of $S$ remains small over the entire temperature range, reflecting the semimetallic band structure, where the electron- and hole-like carriers compensate with each other.
The polar transition is discernible as a clear drop in $S$ upon cooling with a thermal hysteresis, as denoted by arrows.
With increasing $x$, the positive value of $S$ progressively increases, because the hole-like carriers become dominant.
What is prominent is the temperature profile of $S$ for $x=0.08$; the $S$ value unusually increases up to as large as 85 $\mu$V/K with decreasing temperature down to $\sim$50 K, followed by a steep decrease down to zero towards the lowest temperature.
The evolution of $S$ towards low temperatures tends to be weakened as $x$ increases from 0.08 to 0.15.
Consequently, the $x=0.15$ sample shows a peculiar temperature profile of $S$ that is almost constant ($\sim$65 $\mu$V/K) between 60 K and 300 K.
With further increasing $x$ up to 0.22, the value of $S$ decreases in all the temperature range and the temperature dependence becomes monotonic.
%
\par
%
A more suggestive presentation of the thermopower data is the contour plot of $S/T$ as functions of $x$ and $T$ shown in Fig. 1B, since the quantity of $S/T$, relevant to both transport and thermodynamic properties, should be temperature-independent for the conventional Fermi liquid systems (see Eq. \ref{eq:Mott}).
Near the phase boundary between polar orthorhombic and nonpolar monoclinic phases, intriguingly, the $S/T$ magnitude evolves towards the lowest temperature to form a red-colored dome-shaped distribution around $x=0.1$.
In this ``hot'' region, we observed distinct behavior in other physical properties as well, such as a large increase in $\rho_0$ and $\rho_{yx}$, and reduction in SHG intensity, as mentioned above. 
Below we discuss their possible origin in terms of anomalous scattering promoted near the critical regime between the polar and nonpolar phases.
%
\subsection*{Comparison with theoretical calculations.}
%
Figures 4A and 4B present the full landscape of $S$ for 1T'-Mo$_{1-x}$Nb$_x$Te$_2$ obtained by experiments and calculations, respectively.
At high temperatures, the first-principles calculations roughly reproduce the $x$ dependence of the experimental $S$ value, which are featured by the broad hump-like structure around $x=0.1$ - 0.15.
At low temperatures, on the other hand, the calculated results largely deviate from the experimental ones.
The characteristic $S$ hump observed at low temperatures near $x=0.1$ is not reproduced by the calculations; the calculated $S$ value is only one tenth of the measured one below 50 K.
Note here that the electronic specific heat coefficient $\gamma$ is nearly constant ($\gamma\!\sim\!3$ mJ/mol K$^2$ in experiments) as a function of $x$ (inset to Fig. 4B and fig. S3), indicating that the $S$ enhancement observed below 50 K does not originate from an anomaly in band structure, i.e., electronic density of states.
Furthermore, the almost $x$-independent profile of $\gamma$ is theoretically reproduced (inset to Fig. 4B), which affirms the validity of the first-principles calculation for the band structure in the present systems.
(The experimental $\gamma$ values larger than theoretical ones may arise from the renormalization due to the electron-phonon and/or electron-electron interactions.)
To seek the origin of the discrepancy in the $S$ profiles, we now consider the diffusion thermopower based on the Mott formula,
\begin{equation}
S=-\frac{\pi^2}{3}\frac{k_{\rm B}^2 T}{e}\left.\frac{\partial\ln \sigma(\varepsilon)}{\partial\varepsilon}\right|_{\varepsilon=\varepsilon_{\rm F}},\label{eq:Mott}
\end{equation}
where $\sigma$ is the DC electrical conductivity tensor, $k_{\rm B}$ the Boltzmann constant, and $\varepsilon_{\rm F}$ the Fermi energy.
Assuming a simple relation $\sigma\!\sim\!v\tau S_{\rm F}$ with $v$, $\tau$, and $S_{\rm F}$ being the Fermi velocity, (energy dependent) relaxation time, and Fermi surface area, respectively, we obtain
\begin{equation}
S=-\frac{\pi^2}{3}\frac{k_{\rm B}^2 T}{e}\left\{\frac{\partial\ln(vS_{\rm F})}{\partial\varepsilon}+\frac{\partial\ln\tau}{\partial\varepsilon}\right\}_{\varepsilon=\varepsilon_{\rm F}}.\label{eq:S-tau}
\end{equation}
In the present calculation, the first component associated with the band structure is fully taken into consideration, whereas the second one, which stems from the charge relaxation due to scattering, is ignored by adopting the constant $\tau$ approximation.
The deviation of the experimental results from the calculated ones is hence attributable to the dominant contribution from the latter scattering term with strongly energy-dependent $\tau(\varepsilon)$.
%
\par
%
A similar peak structure in the $S$ profile is often observed at low temperatures as a manifestation of a phonon drag effect, which is, however, not the case here.
This is because the polycrystalline samples of doped MoTe$_2$ exhibit essentially the same $S$ profiles as functions of $x$ and $T$, irrespective of substitution species, Nb or Ta (fig. S4).
In general, the phonon drag effect tends to be appreciable at low temperatures in clean systems, where phonons are mostly scattered by electrons via the electron-phonon interaction.
In the polycrystalline samples, since phonons are primarily scattered by grain boundaries and doped impurities, the phonon-drag thermopower would be negligibly small and hence cannot explain the observed $S$ enhancement insensitive to the phonon scattering.
%
\section*{Discussion}
%
It was recently reported that thermoelectric contribution from the second term in Eq. \ref{eq:S-tau} can be dominant in some systems, including heavy fermion compounds\cite{Sun2013PRL,Zlatic2007PRB} and correlated semiconductors\cite{Sun2015NatCom}.
In those materials, the highly dispersive $\tau(\varepsilon)$ is considered to originate from the asymmetric energy-dependent charge relaxation by the local Kondo scattering and multiple relaxation processes.
In the present system, since the dispersive $\tau(\varepsilon)$ seems to evolve near the critical region of the polar-nonpolar structural transition, its origin may be sought in the fluctuations or phase segregation in crystal structure.
For Cu$_2$Se, for instance, a marked enhancement (by $\sim$60\%) of thermopower was observed during a continuous (second-order-like) structural transition, where critical scattering should be induced by fluctuation in crystal structure and density\cite{Liu2013AdvMat}.
Although the polar structural transition in pristine 1T'-MoTe$_2$ is of first-order with minimal fluctuations, the transition is significantly smeared out when it is suppressed to low temperatures by the Nb substitution.
Considering the unusual decrease in SHG intensity below $T_{\rm S}$ for $x=0.08$, furthermore, the polar phase on the verge of the critical composition might be subjected to strong fluctuation or phase separation with the nonpolar one at low temperatures.
This would result in some critical scattering phenomena, causing an anomalous increase in $S$ as well as $\rho_{yx}$ and $\rho_{0}$.
%
%
\par
%
It is noteworthy to mention that the anomalous enhancement in $S$ is absent in the polar metallic state locating away from the critical point.
This is experimentally verified by measuring the thermopower for semimetallic W$_{1-y}$Ta$_y$Te$_2$ that keeps the polar orthorhombic structure in the whole investigated range of temperature and Ta content ($0\!\le\!y\!\le\!0.2$).
We observed no enhancement in $S$ at low temperatures for this series of compounds, resulting in the $S$ landscape similar to that calculated for 1T'-Mo$_{1-x}$Nb$_x$Te$_2$ (see Fig. 4C and fig. S5).
%
\par
%
We here note that a small peak of thermopower is discernible around 30 K even for pristine $x=0$, which is not reproduced by the calculation.
Since the electron and hole-like carriers coexist for $x=0$, even a small change in compensated thermopower by each carrier could result in its sign change and complex temperature variation, which is difficult to reproduce within the constant-$\tau$ approximation.
As another origin of the small peak in thermopower for $x$=0 , we may be able to point out the contribution from the scattering by low-energy phonons activated in the polar phase\cite{Chen2016NanoLett,Zhang2016arxiv}, the energies of which roughly correspond to the temperature range of the thermopower peak.
Due to the multi-carrier effects, however, it is not practical to exactly explain this weak structure in thermopower.
%
\par
%
The origin of the highly dispersive $\tau(\varepsilon)$ is still unclear and needs theoretical supports.
It may however give us a clue that $\tau(\varepsilon)$ significantly differs between the polar and nonpolar phases.
This is made clear by the scaling law on the magnetoresistance effect (so called Kohler's law), which exhibits a substantial change across $T_{\rm S}$ (fig. S6).
The fluctuation or mixing of the phase with a different scattering process might effectively yield a large gradient in $\tau(\varepsilon)$\cite{Sun2015NatCom}.
The critical state of the polar-nonpolar transition thus provides a nice arena for enhancing the impact of polar order on transport properties by achieving the heterogeneous relaxation of charge carriers, which can be a promising key for designing novel thermoelectric materials.
In fact, the $x=0.08$ compound exhibits the peak of thermoelectric figure of merit $Z\!=\! S^2/\rho\kappa\!\sim\! 7.5\!\times\! 10^{-4}$ K$^{-1}$ at $\sim$40 K, where $\kappa$ is the thermal conductivity measured concomitantly with thermopower (see fig. S7, where the dimensionless figure of merit $ZT$ is also plotted).
This is comparable to the $Z$ value for Na$_x$CoO$_2$ known for one of the best hole-type thermoelectric materials ($Z\!\sim\! 1.6\!\times\! 10^{-3}$ K$^{-1}$ at 40 K)\cite{Lee2006NatMater}.
Below 20 K, the $x=0.08$ compound even exceeds Na$_x$CoO$_2$ in $Z$ value, indicating the potential for the thermoelectric applications in the cryogenic region.
\section*{Materials and Methods}
\subsection*{Crystal growth.}
Single crystals of 1T'-Mo$_{1-x}$Nb$_x$Te$_2$ were synthesized with chemical vapor transport technique using iodine as a transport agent\cite{Zandt2007JAlloy}.
Stoichiometric mixtures of Mo (purity 99.9\%), Nb (purity 99.9\%), and Te (purity 99.999\%) powders were sealed in an evacuated quartz tube together with 5 mg/cm$^3$ iodine (purity 99.99\%).
The ampoule was place in a three-zone furnace with a typical temperature gradient from 1050-1000$^{\circ}$C to 920-900$^{\circ}$C.
After 1-2 weeks, we rapidly cooled the ampoule down to room temperature to avoid the formation of the 2H phase\cite{Keum2015NatPhys} and obtained grayish metallic crystals of the metastable 1T' phase with a typical dimension of 2-5 mm $\times$ 0.5-1 mm $\times$ 0.02-0.1 mm, where the longest dimension corresponds to the crystallographic $b$ axis (along the Mo-Mo zigzag chain).
The Nb concentration $x$ was determined by the energy dispersive x-ray microanalysis (EDX), which revealed that the actual $x$ value is slightly smaller than the nominal one (see fig. S1).
We also checked the Raman spectra for the obtained crystals at room temperature, where the peak near 260 cm$^{-1}$ exhibits clear redshift almost linearly with increasing $x$ (inset to fig. S1).
Polycrystalline samples (Mo$_{1-x}$Nb$_x$Te$_2$, Mo$_{1-y}$Ta$_y$Te$_2$, and W$_{1-y}$Ta$_y$Te$_2$) were synthesized by solid state reaction in evacuated quartz tubes\cite{Ikeura2015APLMat}.
The mixtures of stoichiometric amounts of elements in powder form were first heated at 1050-1100$^{\circ}$C for 12 h.
The obtained samples were then ground, pelletized, and annealed at 800-1100$^{\circ}$C for 12 h, followed by water quenching, except for W$_{1-y}$Ta$_y$Te$_2$, the annealing of which in the same manner was followed by furnace cooling.
\subsection*{Transport and thermodynamic measurement.}
Electrical resistivity (along the $b$ axis) and Hall resistivity were measured on the single-crystalline samples by a conventional 5-terminal method with electrodes formed by room-temperature curing silver paste.
The measurements were performed from 2 K to 300 K at the magnetic fields (parallel to the $c$ axis) up to 9 T, using Physical Properties Measurement System (Quantum Design).
The thermopower and thermal conductivity were measured by a conventional four-terminal steady-state method with a temperature difference of less than 1 K (typically 2-4\% of the measurement temperature below 50 K) between the voltage electrodes, using a He closed-cycle refrigerator (from 10 K to 300 K).
Specific heat was measured on the polycrystalline samples from 1.9 K to 10 K with the Physical Properties Measurement System. 
\subsection*{Optical spectroscopy.}
The optical SHG was measured with 1.11 eV fundamental photons (120 fs at 1 kHz) focused on the $c$-plane of the sample (0.2-0.7 mW on $\sim$80 $\mu$m spot) in reflection geometry (see fig. S2A for the setup.)
The incident polarization was controlled by a $\lambda/2$ wave plate, and the polarization of the reflection SHG was analyzed by a Glan laser prism.
The signal was normalized by that of a reference potassium dihydrogen phosphate (KDP) crystal, and accumulated more than 10$^4$ times.
The Raman spectra were measured by a commercial apparatus (RMP-510, JASCO), equipped with a 532 nm laser and a CCD detector.
The measurement was performed on the single crystals in the reflection geometry with the incident and scattered light propagating nearly along the $c$ axis.
\subsection*{Single-crystal x-ray diffraction.}
The diffraction data were collected using a RIGAKU AFC-8 diffractometer equipped with a Saturn70 CCD detector with  Mo$K\alpha$ radiation by an oscillation method.
X-rays were monochromated and focused by a confocal mirror.
The data were measured with the oscillation angle and camera distance of 0.5$^\circ$and 40 mm, respectively.
The initial structure of the low-temperature phase was solved by a direct method using the programs SIR2004\cite{Burla2005JAC}, and refined by a full matrix least-squares method on $F^2$ using the program SHELXL2014\cite{Sheldrick2015Acta}.
\subsection*{Electronic-structure calculation.}
Electronic structure calculations were performed within the context of density functional theory using the Perdew-Burke-Ernzerhof exchange-correlation functional modified by the Becke-Johnson potential, as implemented in the WIEN2K program\cite{WIEN2K}.
Relativistic effects, including spin-orbit coupling, were fully included.
The muffin-tin radius of each atom $R_{\rm MT}$ was chosen such that its product with the maximum modulus of reciprocal vectors $K_{\rm max}$ becomes $R_{\rm MT}K_{\rm max}$ = 7.0.
The structural parameters were taken from the results of single-crystal structural analysis at 100 K for orthorhombic 1T'-MoTe$_2$ obtained in the present study (table S1).
The corresponding Brillouin zone was sampled by a 8 $\times$ 16 $\times$ 4 $k$-mesh.
\par
For the calculation of Seebeck coefficient, we have created an 88-band tight binding model using maximally localized Wannier functions\cite{Souza2001PRB,Mostofi2008CPC,Kunes2010CPC}.
We choose valence $p$-orbitals of Te and $d$-orbitals of Mo as the projection centers of the Wannier functions.
This model is then incorporated into the Boltzmann equation to calculate the Seebeck coefficient using a 50 $\times$ 50 $\times$ 30 $k$-mesh, where we assume a constant relaxation time for all the energy bands
The effects of Nb substitution were treated using the rigid band approximation, where the Fermi level is shifted down to an appropriate energy corresponding to the given hole concentration.
\section*{Supplementary Materials}
Supplementary Material accompanies this paper at {\small {\tt http://www.scienceadvances.org/}}.\\
fig. S1. Nb concentration determined by the EDX microanalysis. \\
fig. S2. Optical setups and SHG characteristics for single-crystalline 1T'-MoTe$_2$. \\
fig. S3. Temperature profile of specific heat for polycrystalline 1T'-Mo$_{1-x}$Nb$_{x}$Te$_{2}$\\
fig. S4. Temperature dependence of resistivity and thermopower for polycrystalline 1T'-Mo$_{1-x}$Nb$_{x}$Te$_{2}$\\
fig. S5. Temperature dependence of resistivity and thermopower for polycrystalline W$_{1-x}$Ta$_{x}$Te$_{2}$\\
fig. S6. Kohler plots for single-crystalline 1T'-Mo$_{1-x}$Nb$_{x}$Te$_{2}$\\
fig. S7. Temperature dependence of thermoelectric figure of merit for single-crystalline 1T'-Mo$_{1-x}$Nb$_{x}$Te$_{2}$\\
fig. S8. Mo-Mo zigzag chain structure in 1T'-MoTe$_2$\\
fig. S9. Calculated thermopower $S$ for monoclinic and orthorhombic 1T'-Mo$_{1-x}$Nb$_{x}$Te$_{2}$\\
fig. S10. Calculated band structures for monoclinic and orthorhombic 1T'-MoTe$_{2}$\\
Table S1. Crystallographic data for 1T'-MoTe$_2$ at 100 K (Orthorhombic)\\
References \cite{LePage1987JAC,Spek2009Acta,Mishina2000PRL,Ogawa2009PRB,Bloembergen1962PR}\\
%
%
%
\bibliographystyle{ScienceAdvances}

%
%
\noindent \textbf{Acknowledgements:} 
The authors are grateful to K. Ishizaka, R. Arita, Y. Nii, and D. Morikawa for fruitful discussions. \\
\noindent \textbf{Funding:} This study was partly supported by the Japan Society for the Promotion of Science (JSPS) Grants-in-Aid for Exploratory Research No. 15K13332, Iketani Science and Technology Foundation, the Murata Science Foundation, Asahi Glass Foundation, and JST PRESTO program (Hyper-nano-space design toward Innovative Functionality).\\
\noindent \textbf{Author Contributions:} S.I. conceived the project. H.S. and S.I. planned and guided the experiments. K.I. performed single crystal growth. K.I. and H.S. measured the transport and thermodynamic properties. M.S.B calculated the band structure and thermopower. N.O. performed the optical SHG measurements. K.I., H.S., and D.H. measured single-crystal x-ray diffraction and D.H. performed the detailed structural analyses. K.I., H.S., and J.F. performed Raman spectroscopy. J.F. and Y.T. contributed to the discussion of the results. H.S. and S.I. wrote the manuscript with contributions from all authors. \\
\noindent \textbf{Competing Interests:} The authors declare that they have no competing financial interests.\\

%
%
\clearpage
%
\noindent {\bf Fig. 1.} \textbf{Tunable polar structural transitions.} ({\bf A}) Lattice structures for 1T'-MoTe$_2$ at 100 K (left) and 300 K (right), deduced from structural analyses based on single-crystal x-ray diffraction. Mo-Mo bonds enclosed by dashed curves form a zigzag chain of Mo atoms along the $b$ axis (see also fig. S8). The Te ions are color-coded depending on the crystallographic sites. Small black arrows near the Te ions schematically denote the displacements relative to the center of Mo ions in the low-temperature orthorhombic phase. ({\bf B}) The contour plot of $S/T$ (thermopower divided by temperature) and structural phase diagram as functions of temperature $T$ and Nb concentration $x$ for 1T'-Mo$_{1-x}$Nb$_x$Te$_{2}$. The cycles and squares are the transition temperatures $T_{\rm S}$ determined by the resistivity and optical second-harmonic generation (SHG), respectively. The open and closed symbols correspond to the cooling and warming runs, respectively. The gray line is a guide to the eyes. Inset displays the temperature variation of single crystal x-ray diffraction images around the 1 0 13 and $\bar 1$ 0 13 reflections in the monoclinic phase for $x$=0, 0.08, and 0.22. These Bragg reflections coalesce into 1 0 13 in the orthorhombic phase. The direction of the horizontal arrow on the photograph corresponds to that of $2 \theta$ angle of the diffractometer.\\
\\
%
\noindent {\bf Fig. 2.} \textbf{Polar metallic nature.} ({\bf A, B}) Temperature dependence of resistivity $\rho$ measured along the $b$ axis (A) and SHG intensity $I_{\rm SHG}$ measured with the incident photon energy at 1.11 eV (B) for 1T'-Mo$_{1-x}$Nb$_{x}$Te$_{2}$ single crystals. The optical SHG signals were measured in the reflection geometry on the $c$-plane, where the incident plane contains the $a$ and $c$ axes. The open and closed circles in (B) denote the warming and cooling runs, respectively.\\
\\
%
\noindent {\bf Fig. 3.} \textbf{Variation of thermopower and Hall resistivity.} ({\bf A}) Temperature profile of thermopower $S$ along the $b$ axis for 1T'-Mo$_{1-x}$Nb$_x$Te$_{2}$ single crystals. ({\bf B}) Magnetic-field profile of Hall resistivity $\rho_{yx}$ at various temperatures for $x$=0 (left), 0.08 (middle), and 0.22 (right). Inset shows the magnetoresistance effect $\Delta \rho(B)/\rho(B=0)=[\rho(B)-\rho(0)]/\rho(0)$ as a function of field at 2 K for $x$ from 0 to 0.22.
\\
\\
%
\noindent {\bf Fig. 4.} \textbf{Comparison between the experimental and theoretical results on thermopower and electronic structures.} ({\bf A, B}) Experimental (A) and theoretical (B) thermopower $S$ as functions of temperature $T$ and Nb concentration $x$ for 1T'-Mo$_{1-x}$Nb$_x$Te$_{2}$. The calculations were performed based on the orthorhombic polar structure. (Quantitatively similar calculation results were obtained for the monoclinic structure, as shown in fig. S9) Inset displays the $x$ dependence of electronic specific heat coefficient $\gamma$, deduced from experiments (closed circles) and from first-principles calculation (open circles). ({\bf C}) Experimental thermopower for polycrystalline W$_{1-y}$Ta$_{y}$Te$_2$ is plotted as functions of temperature and Ta concentration ($y$). ({\bf D}) The calculated Fermi surface of 1T'-Mo$_{1-x}$Nb$_x$Te$_{2}$ with the orthorhombic structure for $x$=0, 0.1, and 0.2, showing electron (blue) and hole (red, green, and magenta) pockets. Since the orthorhombic phase has no inversion symmetry, the Fermi surface sheets are doubled (each corresponding to one spin state). For the difference in band structure between the polar (orthorhombic) and nonpolar (monoclinic) structures, see fig. S10.\\
\\
%
\begin{figure}
\begin{center}
\includegraphics[width=\linewidth]{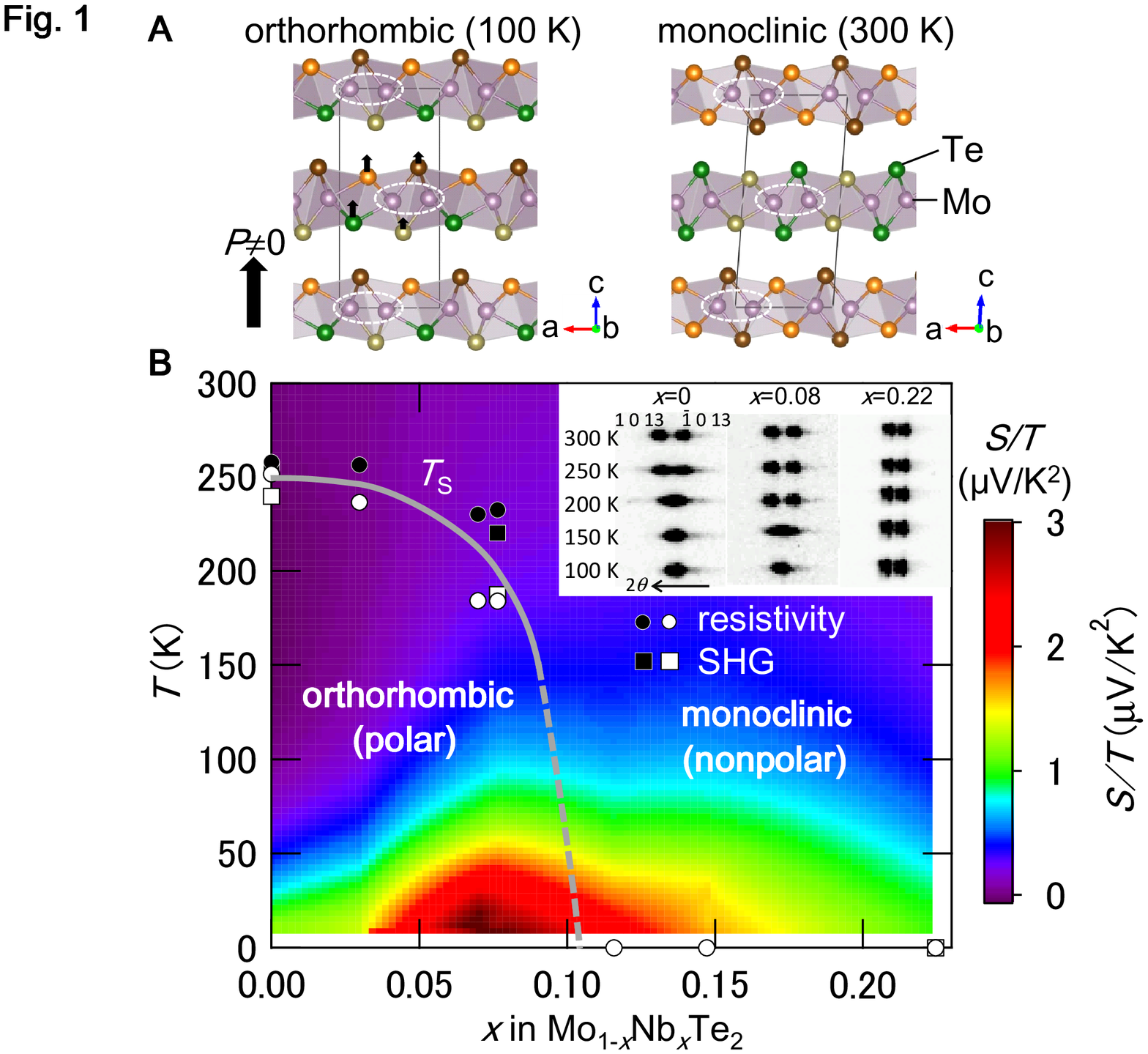}
\end{center}
\end{figure}
%
\begin{figure}
\begin{center}
\includegraphics[width=0.85\linewidth]{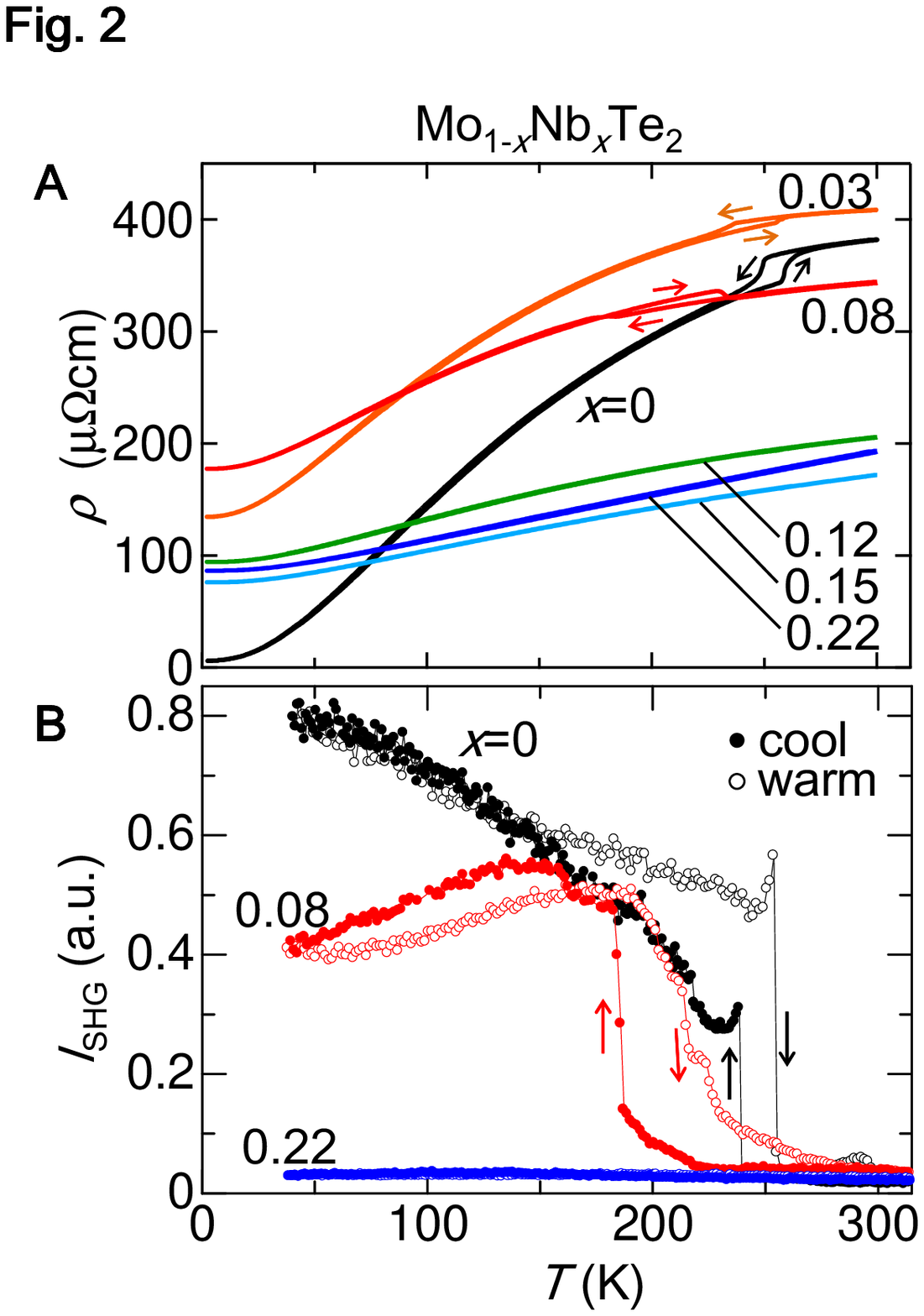}
\end{center}
\end{figure}
%
\begin{figure}
\begin{center}
\includegraphics[width=0.9\linewidth]{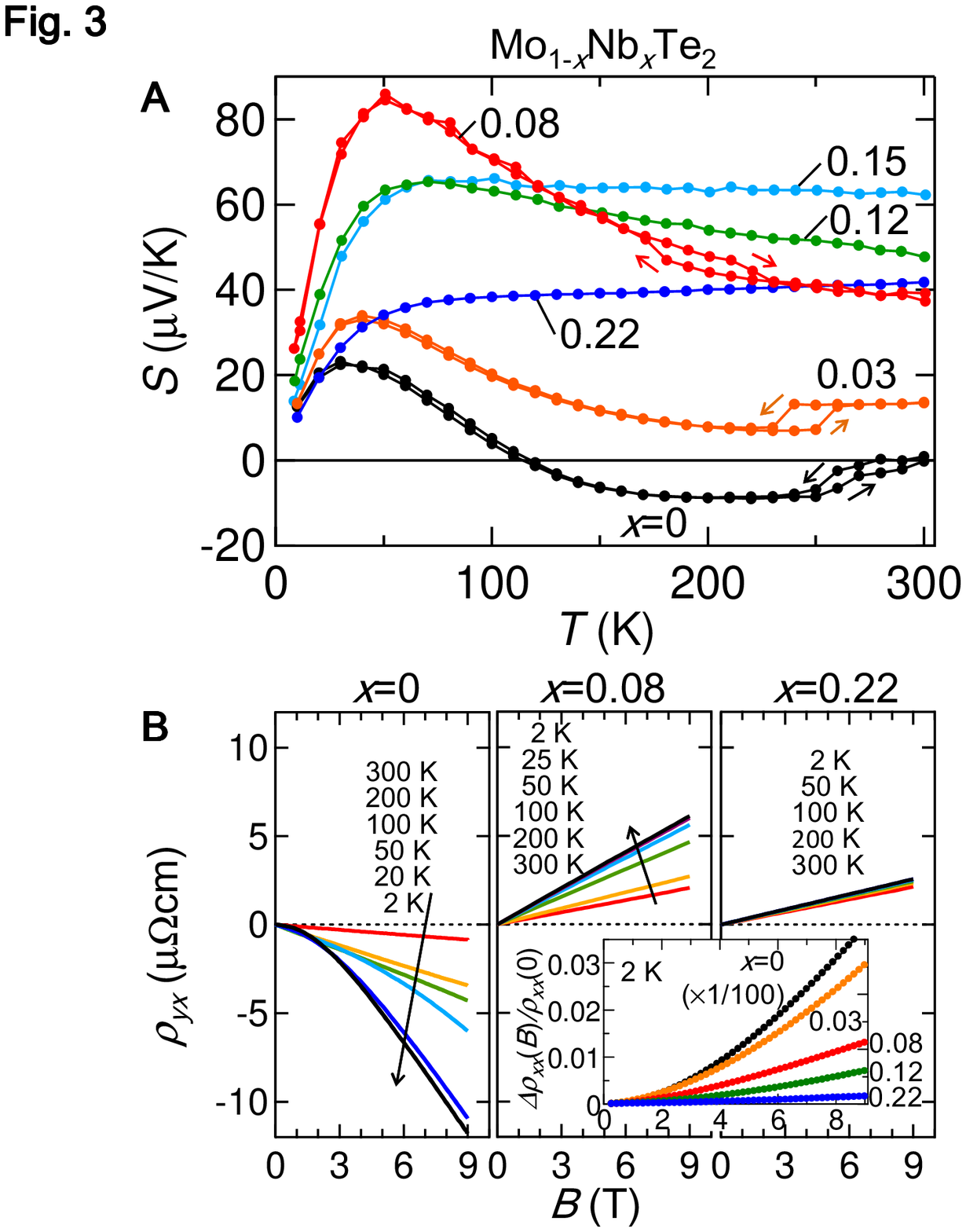}
\end{center}
\end{figure}
%
\begin{figure}
\begin{center}
\includegraphics[width=\linewidth]{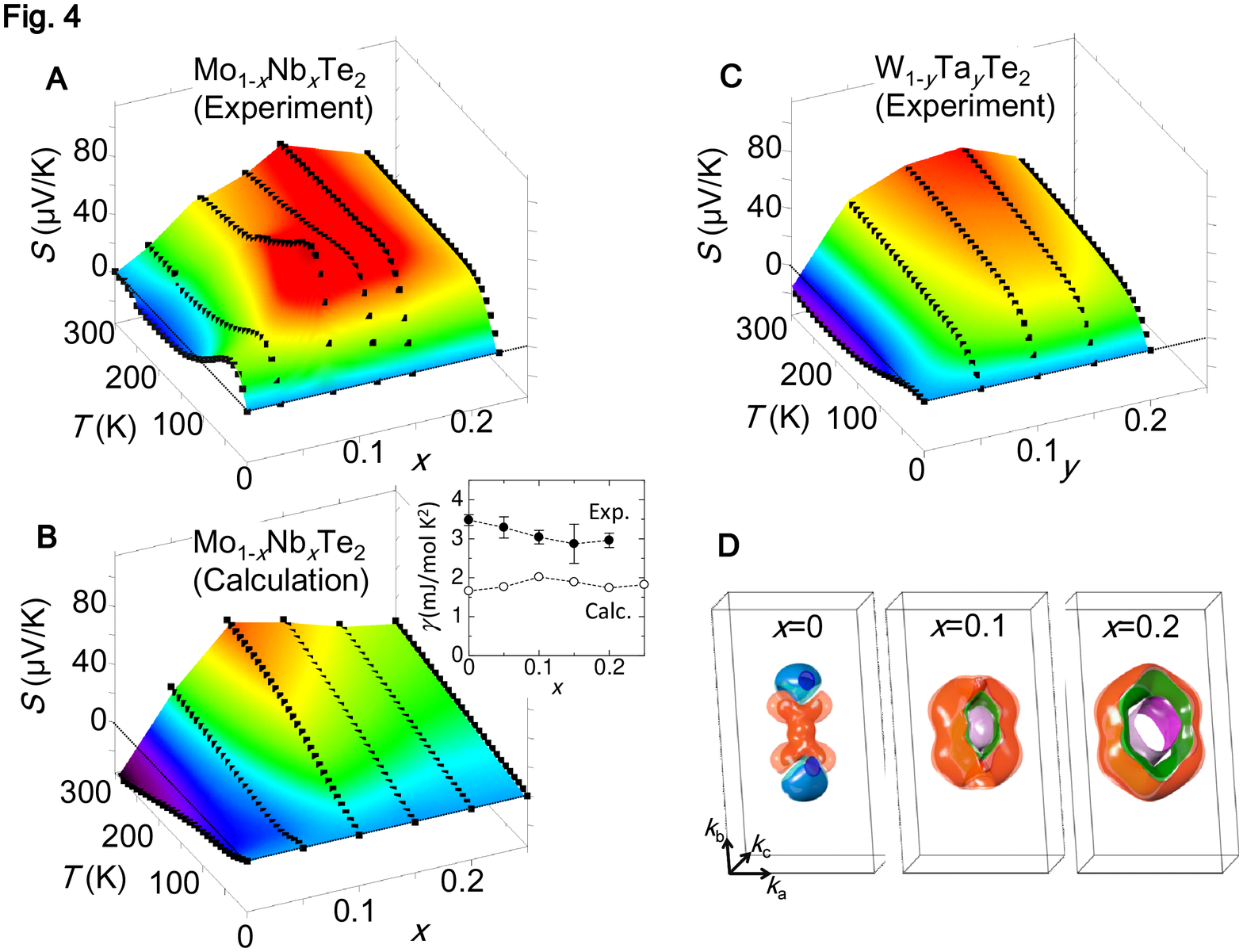}
\end{center}
\end{figure}
%
\end{document}